\documentclass[prl,twocolumn,showpacs,hyperref,balancelastpage]{revtex4}%
\usepackage{epsfig}
\usepackage{color}
\usepackage{ulem}
\usepackage{amsmath}
\usepackage{amsfonts}
\usepackage{amssymb}
\usepackage{graphicx}%
\newcommand{\be}{\begin{equation}}
\newcommand{\ee}{\end{equation}}
\newcommand{\bea}{\begin{eqnarray}}
\newcommand{\eea}{\end{eqnarray}}

\begin{document}
\title{Topological properties of Fibonacci quasicrystals : A scattering  analysis of Chern numbers}
\author{E. Levy\textsuperscript{1,2}, A. Barak\textsuperscript{2},
A. Fisher\textsuperscript{1}, and E. Akkermans\textsuperscript{1}}
\affiliation{$^1$Department of Physics, Technion Israel Institute of Technology, Haifa 32000, Israel}
\affiliation{ $^2$Rafael Ltd., P.O. Box 2250, Haifa 32100, Israel}

\begin{abstract}

We report on a study of topological properties of Fibonacci quasicrystals.  Chern numbers which label the dense set of spectral gaps, are shown to be related to the underlying palindromic symmetry. 
Topological and spectral features are related to the  two independent phases of the scattering matrix: the total phase shift describing the frequency spectrum and the chiral phase sensitive to topological features. Conveniently designed gap modes with  spectral properties directly related to the Chern numbers allow to scan these phases. An effective topological Fabry-Perot cavity is presented.

\end{abstract}

\pacs{41.20.Jb,77.84.Lf,61.44.Br,03.65.Vf,11.55.-m,03.65.Nk,07.60.Ly,78.67.-n}

\date{\today}
\maketitle

Topological features of quasi-periodic (QP) structures have been studied under  various guises both in physics and mathematics \cite{gaplabel,general}. For instance,  Chern integers label the dense set of gaps in the energy (frequency) spectrum. More generally, Chern numbers are known to play a role in  problems where the underlying topology has been identified. Examples are provided by the quantum Hall effect, anomalies in Dirac systems and topological insulators. For each case, the origin of topological features is identified e.g., magnetic field, chirality or nontrivial band structures. However, in QP structures like the Fibonacci chain, the physical origin underlying the existence of Chern numbers has not yet been identified. The purpose of this letter is to address this point  and to propose new expressions and possible measurements of Chern numbers using scattering data. We shall discuss the specific case of Fibonacci chains, but we expect our results to apply to a larger class of similar QP one-(or higher) dimensional structures. 

\begin{figure}
\includegraphics[viewport=-5bp 16bp 988bp 741bp, clip,width=1\columnwidth]{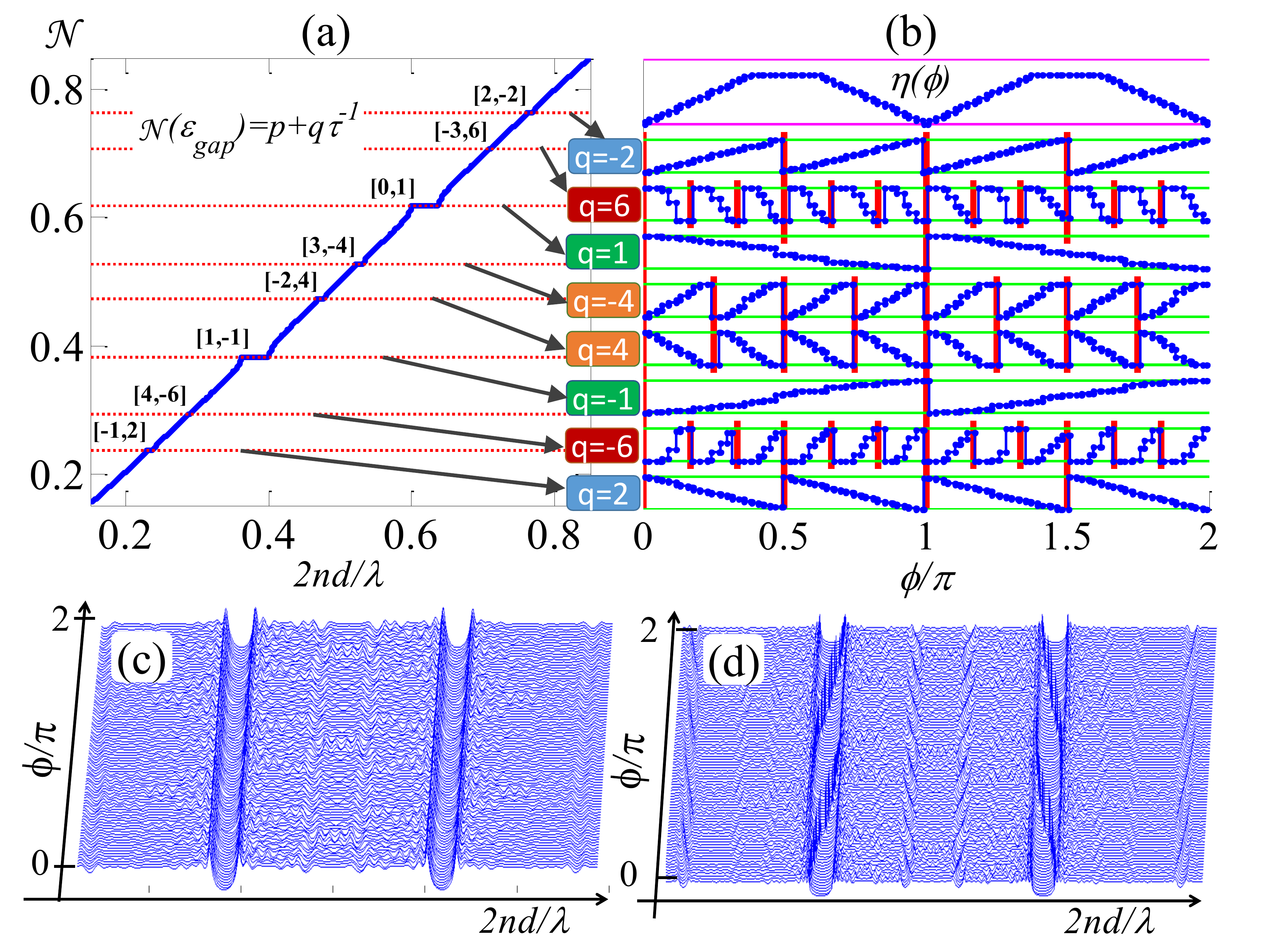}

\caption{(color online) (a) Normalised IDOS for the Fibonacci chain  \cite{rkstructure}. The location of 8 selected gaps using their topological numbers $[p, q]$ are indicated in accordance to (\ref{eq:fibogaptopology}). (b) The corresponding  $\protect\overrightarrow{F}_N\protect\overleftarrow{F}_N$  gap modes behaviour  as a function of $\phi$ compared to $\eta (\phi)$ (top). (c)-(d) $\rho(k)$, obtained from (\ref{theequation}) for $\protect\overrightarrow{F}_N$ (c), and $\protect\overrightarrow{F}_N\protect\overleftarrow{F}_N$ (d).\label{spectral}}
\end{figure}

QP systems share similar spectral properties \cite{gaplabel,general,general2, hof, aubry,damanik,spontem}, e.g. their singular continuous frequency spectrum has a fractal structure (Cantor set) \cite{reviewfractals}, with a dense distribution of gaps characterised by the gap labelling theorem \cite{gaplabel}. It provides a precise expression of the (normalised to unity) integrated density of states (IDOS) $\mathcal{N}\left(\varepsilon_{gap}\right)$ at energies $\varepsilon_{gap}$ inside the gaps. For the Fibonacci chain, gaps are labeled by means of two integers $\left( p,q\right) $ such that,
\begin{equation}
\mathcal{N}\left(\varepsilon_{gap}\right)=p+q\tau^{-1},\label{eq:fibogaptopology}
\end{equation}
where $\tau = (1 + \sqrt{5} )/2$ is the golden mean. The  integers $q$ are  Chern numbers \cite{general} and $p(q)$ keeps  $\mathcal{N}\left(\varepsilon_{gap}\right)$  within $[0,1]$. These features which have been widely studied \cite{gaplabel,general,general2,kunz,damanik} and recently measured using cavity polaritons \cite{polaritons}, are shown in Fig.\ref{spectral}.a. 

Here, we consider finite Fibonacci chains and study the behaviour of conveniently emulated edge states (gap modes) driven by an angular variable $\phi$ (defined hereafter)  accounting for the important but not yet  studied palindromic or mirror  symmetry of quasicrystalline chains. As we shall see, $\phi$  whose role has been studied in the Harper model or quantum Hall related setups \cite{hof, aubry,krauss2,dana}, allows to monitor this structural symmetry.

To better understand the meaning of $\phi$, it is instructive to review basic rules used to generate finite QP structures. We study, for simplicity, sequences built using a two letters alphabet $\{A,B\}$. These letters may describe different physical setups such as electromagnetic waves in a dielectric with refractive index modulation or a Schr\"odinger equation with a modulated potential \cite{polaritons}. We first consider the substitution  algorithm (equivalent to the concatenation rules) \cite{substitution,grimm}. A substitution $\sigma$ acts onto $\{A,B\}$ according to $ \sigma (A) = AB$ and $\sigma (B) = A$. Successive applications of $\sigma$ generate a sequence $S_j = \sigma^j (B)$ whose length is the Fibonacci number $F_{j>1} = F_{j-1} + F_{j-2}$, with $F_0\! =\! F_1\! =\!1$. The ratio $F_{j+1} / F_j $ tends to the golden mean $\tau$ in the limit $j \rightarrow \infty$. The corresponding sequence $S_\infty$ becomes rigorously QP and invariant, i.e., self-similar under this iteration transformation. The gap labeling relation (\ref{eq:fibogaptopology}) is established using this algorithm \cite{gaplabel} (see \cite{supplement} for an elementary derivation). A second set of rules includes two alternative approaches. The first \cite{gaplabel,luck}, relies on a characteristic function $\chi_n$ taking two possible values $\pm 1$ respectively identified to the letters $ \{ B,A \}$. Among possible choices and using a discrete analogue of the continuous Aubry-Andre model \cite{aubry}, we consider the form,
\be 
\chi_m = sign \, \left[ \cos \left( 2 \pi \, m \, \tau^{-1} + \tilde\phi \right) - \, \cos \left( \pi \, \tau^{-1} \right) \right], 
\label{chi}
\ee
proposed in \cite{krauss}. The angular degree of freedom $\tilde \phi$ can be safely ignored for $S_\infty$ but not, as we shall see, for the finite segment $\overrightarrow{F}_N \!\!\equiv\!\! \left[ \chi_1 \, \!\chi_2 \,\! \cdots \! \chi_N \right]$ of $S_\infty$. The chain $S_j$ generated using substitution is reproduced by setting $ \tilde\phi \!\!=\! \phi_F \!\equiv \!\pi \tau^{-1}$, and $N\!\!=\!F_j$. We thus redefine (\ref{chi}) using a shifted phase $ \phi \!\equiv\! \tilde\phi \!-\! \phi_F$. 

\begin{figure}
\includegraphics[bb=0bp 22bp 790bp 692bp,clip,width=1\columnwidth]{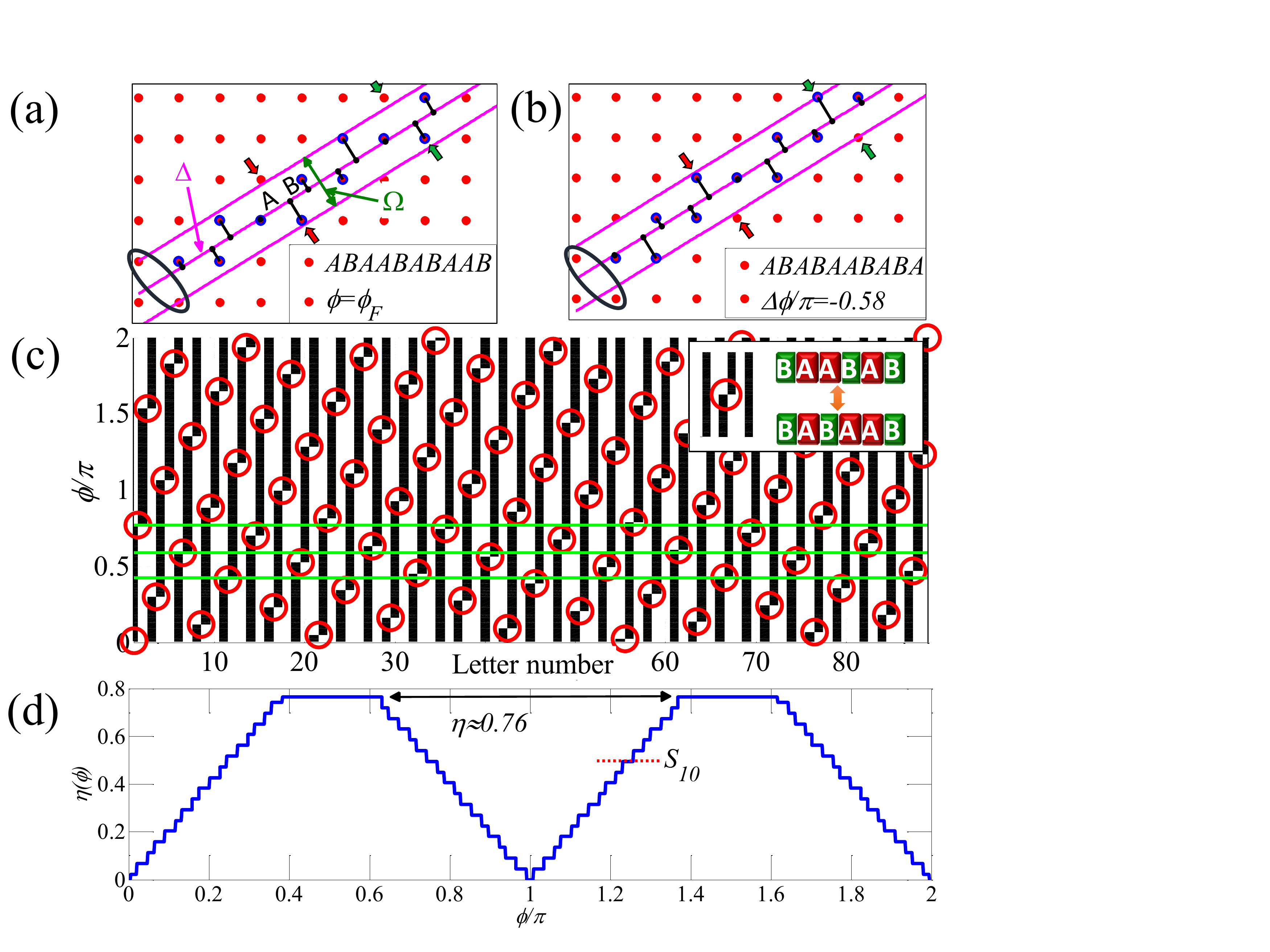}

\caption{(color online) (a)-(b) The C$\&$P method for $\phi_F$ and $\phi_F\!-\!0.58\pi$. An ellipse indicates  the origin of the chain. The resulting $N=10$ sequences are in the insets. Shifting the band $\Omega$ along the vertical axis induces 2 structural changes corresponding to points entering and leaving $\Omega$ (red/green arrows). (c)-(d) Structure properties of  $\protect\overrightarrow{F}_N$  \cite{rkstructure} as a function of $\phi$. (c) Structural color plot. Black(white) regions describe $B(A)$ letters. Identical structural changes (red circles, inset) occur when scanning $\phi$ (e.g. green lines). (d) Behaviour of $\eta \left( \phi \right)$ ($S_{10}$ is indicated). The saturation at $\eta\! \simeq\!0.76$ is specific to Fibonacci through the $[AA]$ pair occurrence, $2 \tau\! -\!3$ \cite{supplement}.  \label{structural}}
\end{figure}

Finally, the Cut$\&$Project method (hereafter C$\&$P) is used to investigate quasicrystals, tilings and dynamical systems \cite{substitution,grimm} (see Fig.\ref{structural}). A QP chain is obtained from the $ \mathbb{Z}^2$-lattice cut by a line $ \Delta $ defined by $y \!= \!\tau^{-1} x \!+\! const $. We denote $ \Delta_\perp $ the direction perpendicular to $\Delta $ and define an acceptance window by the band of width $\Omega$, centered at $ \Delta $. This realises the "cut". A C$\&$P set is obtained by projecting the  $\mathbb{Z}^2$ points inside $\Omega$ on  $ \Delta $ and along $ \Delta_\perp$. The two possible distances along $ \Delta $ between neighbouring projections, are  denoted $\{A,B\}$. We thus generate $S_\infty$ for which the choice of origin on $ \Delta $ is irrelevant.  For finite chains it is not so, since the origin fixes the first letter and the iteration of the sequence. We note that C$\&$P and characteristic function methods are related through the constant term in the equation for $ \Delta $, namely, $y \!= \!\tau^{-1} x \!-\! {\phi \over 2 \pi}$. This allows for a useful interpretation of $\phi$ which accounts for the rearrangement of letters in a Fibonacci chain resulting from its translation along the $y$-axis (Fig.\ref{structural}.b). The $2 \pi$-periodic phase $\phi$ is a structural degree of freedom (a.k.a a phason). For a chain of length $N$, each value of $\phi$ generates a different segment of $S_\infty$, and corresponds to a translation $\delta n \!= \!\left( \tau / 2 \pi \right) \phi $ along $\Delta$. Monitoring $\phi$ induces a series of $N$ identical local structural changes equivalent to the inversion of a single 6-letters segment $BAABAB \!\leftrightarrow\! BABAAB$ (Figs.\ref{structural}.a-b). These changes occur one at a time and are distributed according to a geometrical pattern displayed in Fig.\ref{structural}.c. This discreteness underlies the  characteristic stepwise shapes visible in Figs.\ref{spectral}.b, \ref{structural}.d and \ref{chern11}. 

The combined description using $\phi$ and its associated C$\&$P interpretation allows to unveil an essential structural symmetry of QP chains, namely their palindromic character (mirror symmetry). Sweeping $\phi$ over a period, we observe that any chain of length $N$ becomes perfectly symmetric for two specific values, $ \phi_{pal} (N) \!= \!- (N+2)\pi\tau^{-1} \!+\! m\pi$, where $m\! \in \!\mathbb{Z}$. Deviations from this palindromic symmetry is described using a structural parameter defined in $[0,1]$ by
\be
\eta ( \phi ) \equiv {1 \over N} \sum_{j=0}^{[ {N-1 \over 2} ]} | \chi_{j+1} (  \phi ) - \chi_{N-j} ( \phi) | \, \, .
\label{eta}
\ee
The palindromic cycle for $\eta (  \phi )$ is displayed in Fig.\ref{structural}d. 
This periodic occurrence of palindromic chains, as we shall see, is underlying the existence of Chern numbers in the gap labeling relation (\ref{eq:fibogaptopology}). It is characteristic of C$\&$P QP systems and of the corresponding  Aubry-Andre-Harper tight binding model \cite{hof,aubry}. The structural phase $\phi$ drives a 1D structural symmetry cycle, and thus plays a role analogous to the magnetic field in the 2D quantum Hall effect.
It is therefore natural to set the origin of $\phi$ at the palindromic symmetry, i.e., to redefine $\phi$ as $\phi \!- \!\phi_{pal} (N)$ in (\ref{chi}).
The observation of a $\phi$-driven palindromic cycle is one of the main result of this letter. 

We now introduce the scattering formalism offering an elegant framework to study the spectral consequences of this structural symmetry \cite{meidan}. We stress that although self-similarity and gap labeling (\ref{eq:fibogaptopology}) are spectral properties of the infinite $S_\infty$ (where $ \phi$ is irrelevant), they are equally relevant to finite chains (Fig.\ref{spectral}.c). 
We recall that any quantum or wave system with a potential defined w.r.t a free part can be probed using scattering waves of wave vector $k$. Here, we consider a Fibonacci dielectric medium with scattered electromagnetic waves \cite{supplement}. 
\begin{figure}
\includegraphics[bb=-25bp 38bp 1065bp 400bp,clip,width=1\columnwidth]{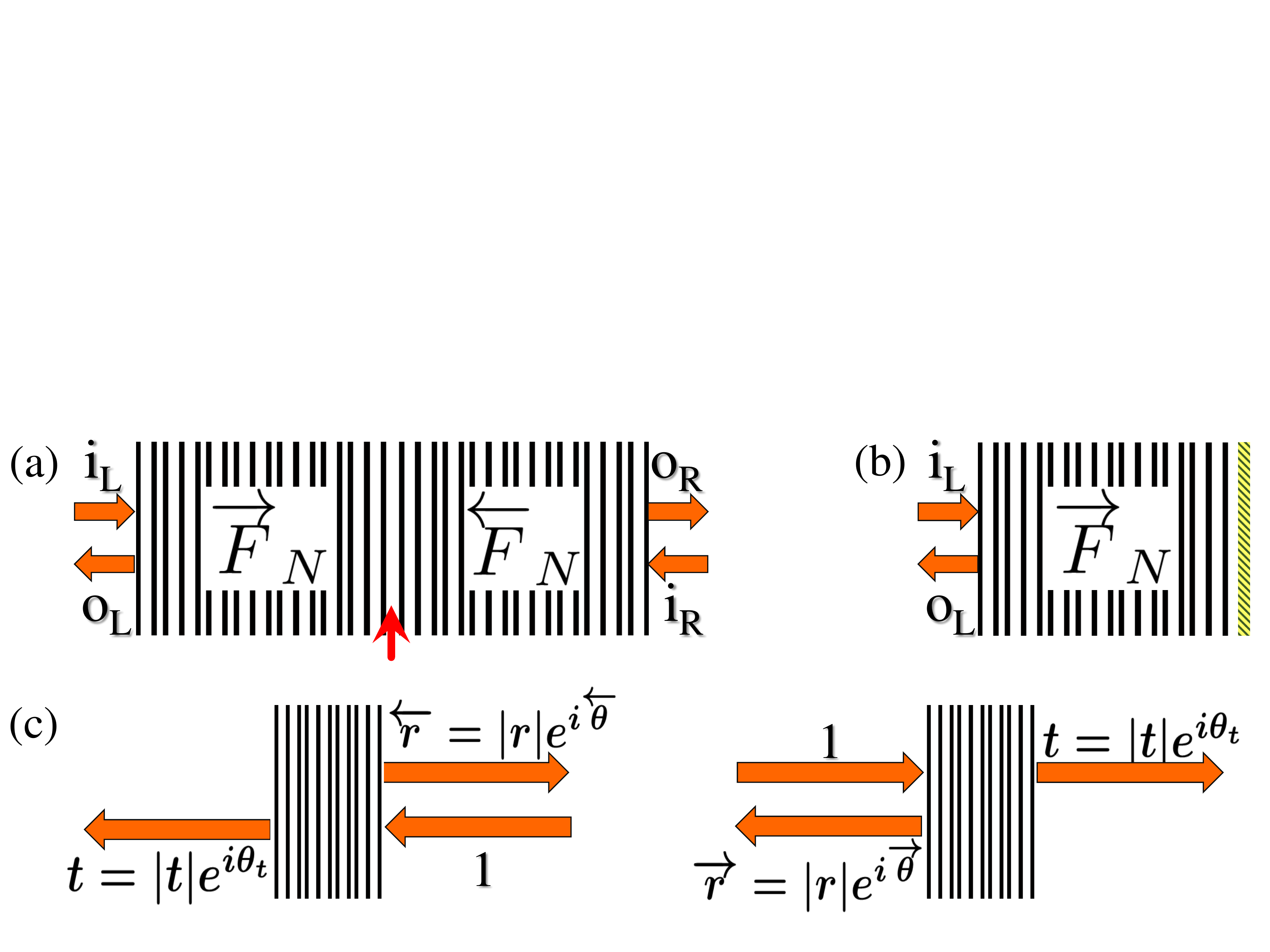}

\caption{(color online) (a) The scattering setup. A finite chain is inserted in between perfect dielectric media supporting incoming $( i_L , i_R )$ and outgoing waves $(o_L , o_R )$. Displayed case: $P \!\equiv \!\protect\overrightarrow{F}_N \!\protect\overleftarrow{F}_N$ ($\uparrow$ indicates the interface) (b) The chain $\protect\overrightarrow{F}_N$ with a reflecting boundary condition (golden bar), is equivalent to the unfolded open structure $P$ in (a). (c) Complex amplitudes corresponding to a wave incident respectively on the right or left side of the chain. 
\label{scattering}}
\end{figure}
With clear enough notations (Fig.\ref{scattering}.a), the (unitary) scattering matrix $S_{\overrightarrow{F}}$ of the sequence $\overrightarrow{F}_N$ is
\begin{eqnarray}
 \left(
\begin{array}{l}
o_L  \\
o_R
\end{array}
 \right)
 =  \left(
\begin{array}{ll}
\overrightarrow{r} \, & t \,  \\
t \,  & \overleftarrow{r} \,
\end{array}
 \right)  \left(
\begin{array}{l}
i_L  \\
i_R
\end{array}
 \right)
 \equiv S_{\overrightarrow{F}}\,
 \left(
\begin{array}{l}
i_L  \\
i_R
\end{array}
 \right) \, .
\label{smatrix0}
\end{eqnarray}
It is diagonalisable under the form $diag ( e^{i \phi_1} , \, e^{i \phi_2})$. We define the total phase shift $\delta (k)\! \equiv \!(\phi_1 (k) + \phi_2 (k) )/2$. The transmission and reflexion amplitudes, $t\!\equiv \!|t|e^{i \, \theta_t}$,  $\overrightarrow{r}\!\equiv \!|r| e^{i \overrightarrow{\theta}_{\overrightarrow{F}}}$, and $\overleftarrow{r} \!\equiv\! |r| e^{i \overleftarrow{\theta}_{\overrightarrow{F}}}$ corresponding to left or right incoming waves are represented on Fig.\ref{scattering}.c. The relation, $\mbox{det}\, S_{\overrightarrow{F}} (k)\!=\! e^{2 i \delta (k)}\! = \!- t / t^*$, implies that $\delta(k)\!=\!\theta_t (k) \!+\!\pi/2$. A simple relation exists between $\delta(k)$ and the density of modes $\rho(k)$ \cite{dunnelevyea}. For $\overrightarrow{F}_N$, it reads  
 \be
\rho (k) - \rho_0 (k) = {1 \over 2 \pi}\, \mbox{Im} \,\frac{\partial}{\partial k} \ln \mbox{det}\, S_{\overrightarrow{F}}(k) = {1 \over \pi} {d \delta (k) \over d k} \, , 
\label{theequation}
\ee
where $\rho_0(k)$ is the density of modes of the free system, i.e. withuot  $\overrightarrow{F}_N$. This relation is used to obtain the spectrum of Fig.\ref{spectral}.c. Since $\rho(k)$ is independent of the incident wave direction, the  phases $\delta$ and $\theta_t$ do not discriminate between the sequence $\overrightarrow{F}_N$ and its reversed $\overleftarrow{F}_N \!\!\equiv \!\!\left[ \chi_N \, \chi_{N-1}\! \, \!\cdots \!\chi_1 \right]$, and therefore they are not sensitive to the symmetry of the sequence (palindromic or not). This insensitivity, in addition to the fact that shifts in $\phi$ are only shifts along $S_\infty$, clarifies why these phases, describing bulk properties, are independent of $\phi$ \cite{rkphidep}. This is not so for the reflexion amplitudes, related by $ \overleftarrow{r} \!= \! \overrightarrow{r} \, e^{i \alpha}$, where $\alpha \!\equiv \!\overleftarrow{\theta}_{\overrightarrow{F}} \!-\! \overrightarrow{\theta}_{\overrightarrow{F}}$ is a chiral phase which, by definition, vanishes identically for a palindromic $\overrightarrow{F}_N$ and is finite otherwise. Within the gaps $|\alpha\left(\phi,k\right)|$ is a function of $\phi$, essentially insensitive to $k$ (Figs.\ref{chern11}a,e), thus making it a spectral counterpart of $\eta (\phi)$. Finally,  the unitarity condition, $t \! \overrightarrow{r}^*\! + \!\overleftarrow{r} \! t^*\!=\!0$, of $S_{\overrightarrow{F}}$ in (\ref{smatrix0}), provides the relation, 
\be
2 \theta_t + \pi  = \overrightarrow{\theta}_{\overrightarrow{F}} + \overleftarrow{\theta}_{\overrightarrow{F}} = 2 \delta \,,
\label{phasesS}
\ee
showing that while the {\it difference} of the two reflected phases is the $\phi$-dependent chiral phase $\alpha$, their {\it sum} is the $\phi$-independent total phase shift $\delta$.
\begin{figure*}
\includegraphics[bb=11bp 5bp 1225bp 325bp,clip,width=2\columnwidth]{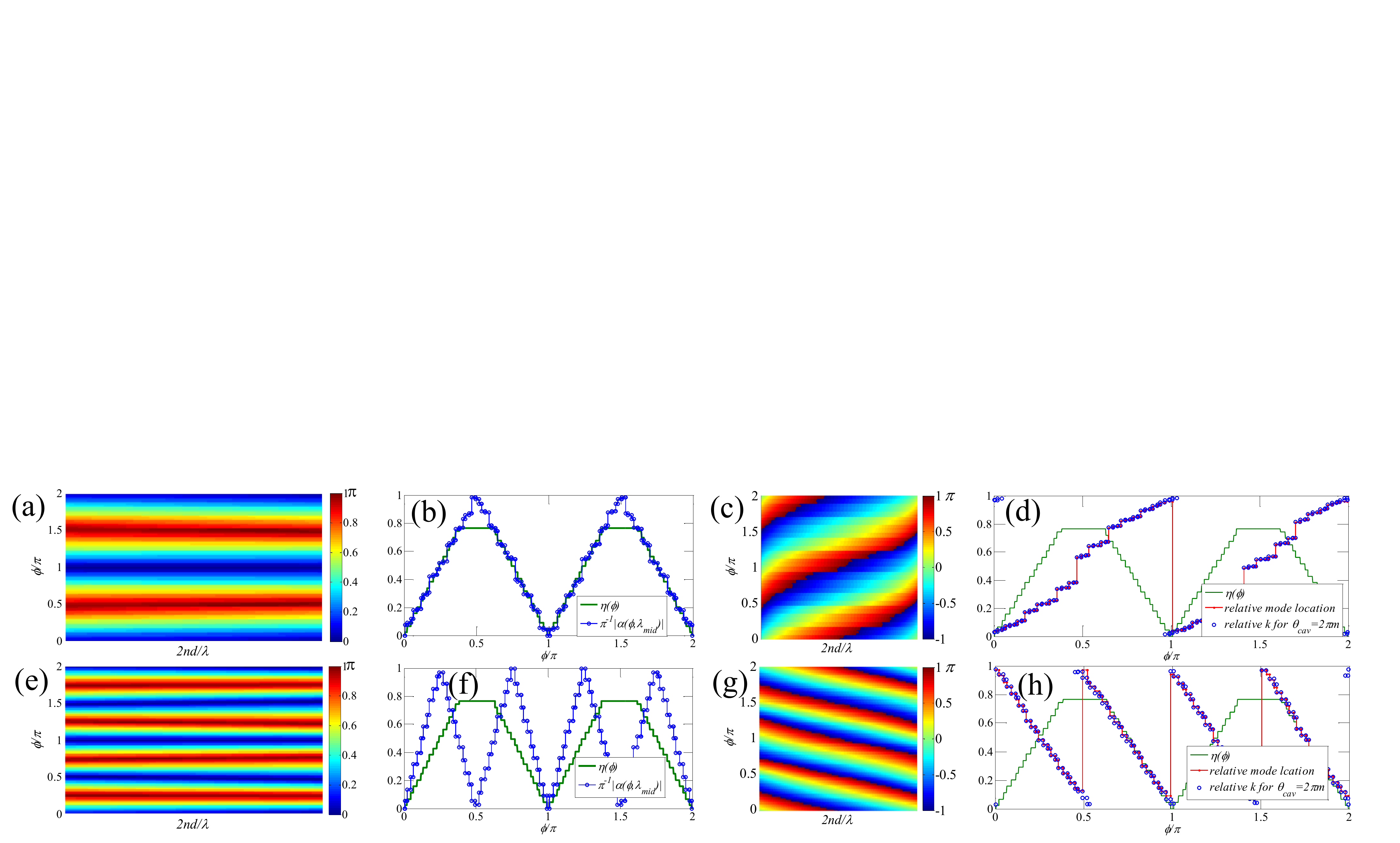}
\caption{(color online) The phases $\alpha_q$ and $\theta_{cav}$ for two gaps of  \cite{rkstructure}, with $q \!= \!-1,2$.  (a) Color plot of $| \alpha_{-1} (\phi,k )|$ within the gap. (b) Comparison between the midgap value, $|\alpha_{-1} (\phi )|$, and $\eta (\phi)$. $| \alpha_q (\phi ) |$ closely follows the palindromic cycle, and is a $\pi / |q|$-periodic spectral counterpart of the $\pi$-periodic $\eta(\phi)$ \cite{rkspecpal} (c) Color plot of $\theta_{cav}(\phi,k )$ within the gap $q \!= \!-1$. Light green areas correspond to the  resonance condition $\theta_{cav}\! = \!2 \pi m$. (d) Crossover of the gap mode $k_m(\phi)$. Relative spectral location (red) is compared to the resonance condition (blue) and to $\eta (\phi)$ (green). (e)-(h) Similar plots for $q=2$. \label{chern11}}
\end{figure*}

The phase $\alpha(\phi)$ is most conveniently investigated using states other than $\phi$-independent scattering states. We thus consider distinct choices of boundary conditions and study the resulting edge states whose spectral locations depend on these choices, e.g. closed (mirror) boundary condition, interface or local defect \cite{rkcredit}. Since each choice leads to edge states of the same topological content \cite{unpub}, without loss of generality, we consider a finite Fibonacci chain (of length $N=F_j$, to properly fix the origin of $\phi$) with reflecting boundary condition at one end (Fig.\ref{scattering}.b).
To adapt this setup to the scattering approach, we unfold it as displayed in Fig.\ref{scattering}.a, so that the resulting artificial palindrome $P\! \equiv \!\overrightarrow{F}_N \overleftarrow{F}_N$ is an open scattering system of length $2N$. Since $P$ is not a {\it bona fide} Fibonacci chain, additional interface modes appear in the gaps of $\overrightarrow{F}_N$ at values $k_m (\phi)$ whose behaviour depends solely on the Chern number $q$ of each gap as assigned by relation (\ref{eq:fibogaptopology}). Modes cross the gaps in a direction set by $sign(q)$ and a crossing period of $\pi / |q|$, as shown in Figs.\ref{spectral}.b,d \cite{rkdoubling}. When $\overrightarrow{F}_N$ is palindromic ($\overrightarrow{F}_N \!=\! \overleftarrow{F}_N$), $P$ reduces to $\overrightarrow{F}_N$ with periodic boundary conditions instead of a mirror. Therefore, no gap modes exist (Fig.\ref{spectral}.b,d). 
 
 These observations become more quantitative by expressing the scattering data of the open structure $P$  in terms of those of  $\overrightarrow{F}_N$. The respective scattering matrices $S_{P,\overrightarrow{F},\overleftarrow{F}}$ of the chains $P$, $\overrightarrow{F}_N$ and $\overleftarrow{F}_N$, are related by,
\be
\mbox{det} \, S_P = \left( \mbox{det}\, S_{\overrightarrow{F}} \right) \, \left( \mbox{det} S_{\overleftarrow{F}} \right) \, e^{i \varphi} \, ,  
\label{SP}
\ee
where $ e^{i \varphi}\!\equiv\! \left( 1\! -\! z^* \right)\! / \!\left(z\! - \!1 \right)$ is a Fabry-Perot factor with $z \!\equiv\! \overrightarrow{r}_{\overleftarrow{F}} \, \, \overleftarrow{r}_{\overrightarrow{F}} \!= \!|r|^2 \, e^{i( \overrightarrow{\theta}_{\overleftarrow{F}}\!+\!\overleftarrow{\theta}_{\overrightarrow{F}}}) \!\equiv \!|r|^2 \, e^{i \theta_{cav}}$ \cite{supplement}. Since by construction, $\overrightarrow{\theta}_{\overleftarrow{F}}\! =\! \overleftarrow{\theta}_{\overrightarrow{F}}$, the cavity phase $\theta_{cav}$, a denomination we shall justify later on, becomes
\be
\theta_{cav} \equiv \overrightarrow{\theta}_{\overleftarrow{F}} + \overleftarrow{\theta}_{\overrightarrow{F}} =  2 \overleftarrow{\theta}_{\overrightarrow{F}} \, .
\label{thetacav}
\ee
Moreover, inserting the relation $\mbox{det}\, S_{\overrightarrow{F}} \!= \!\mbox{det} \, S_{\overleftarrow{F}} \!= \!e^{2 i \delta}$ into (\ref{SP}), leads to $\delta_P = 2 \delta + \varphi / 2 $, $\delta_P$ being the phase shift of $S_P$. A condition for the appearance of a gap mode $k_m$ in the spectrum of $P$ is $ \delta_P (k_m )\!- \!2 \delta (k_m )\!=\! \pi m$, where $ m \!\in \!\mathbb{Z}$ or equivalently $\varphi \! =\! 2 \pi m$ \cite{rklevinson}. 
In the limit $|r|^2 \!\rightarrow\! 1 $ \cite{supplement}, we obtain $ e^{i \varphi}  \!= \!e^{- i \theta_{cav}}$ for the Fabry-Perot factor, so that the resonance condition on $\varphi$ becomes $\theta_{cav} \left( k_m \right)\! =\! 2 \pi m$. Since, according to (\ref{thetacav}), $\theta_{cav}$ is a phase of $S_{\overrightarrow{F}}$, the gap modes $k_m (\phi)$ of $\overrightarrow{F}_N$ bounded by a mirror at one end, can be obtained from the scattering data of the corresponding open chain \cite{rkdoubling}. Knowing $\theta_{cav} \left(  \phi, q, k \right)$ allows to determine precisely the spectral location $k_m (\phi)$ of the gap modes during a palindromic cycle and to characterise their crossing direction and periodicity as a function of the corresponding Chern number $q$ (Figs.\ref{chern11}.c,d,g,h). This quantitative description is the basis of the qualitative analysis displayed in Figs.\ref{spectral}.b,d. Using (\ref{phasesS}) and the definition of the chiral phase $\alpha$ we obtain, 
\be
 \theta_{cav} \left(  \phi, q, k \right) = 2 \, \delta (k) + \alpha_q  \left(  \phi \right), 
 \label{phases}
 \ee
which disentangles  $\theta_{cav}$ into $\delta$, insensitive to $\phi$, and a $k$-independent $\alpha_q$ \cite{rkalphak}, which displays the topological properties, namely the $\pi/|q|$ period in $\phi$. These properties do not depend on the structural modulation strength \cite{supplement}.

As a straightforward consequence of the definition of $\theta_{cav}$, its winding number
\be
{\cal W} \left( \theta_{cav} \right) \equiv {1 \over 2 \pi} \int_0^{2 \pi} d \phi \, {d \theta_{cav} \left(\phi , q , k_m \right)  \over d \phi} = 2 \, q \, \, ,
\label{winding}
\ee
defined in a piecewise manner, provides a direct determination of the corresponding Chern number $q$ \cite{rkhafezi}. In addition, the $\phi$-independence  of $\delta$, together with (\ref{phases}), lead to ${\cal W} \left( \alpha_q \right) \!= \!{\cal W} \left( \theta_{cav} \right)\! =\! 2 q$  for the chiral phase winding number. These expressions are especially interesting since they relate the Chern numbers labeling the spectral gaps  of the infinite chain $S_\infty$ to the scattering matrix of finite chains. It is thus possible to directly deduce Chern numbers from a scattering experiment. 

Finally, we provide another interpretation of the resonance condition $\theta_{cav} \!\left( k_m \right) \!\!=\!\! 2 \pi m$. The interface between $\overrightarrow{F}_N$ and $\overleftarrow{F}_N$ defines a zero length effective cavity between two topological ($q$-dependent) and frequency-dependent highly reflective mirrors. The accumulated phase shift of this effective cavity is precisely the cavity phase $\theta_{cav}\! \!=\! \overrightarrow{\theta}_{\overleftarrow{F}} \!+ \!\overleftarrow{\theta}_{\overrightarrow{F}} $ (Fig.\ref{scattering}.c), thus justifying its name. We define a frequency-dependent effective  cavity length,
\be
{\cal L} \left(  \phi, q, k \right) \equiv  {\lambda (k) \over 4} \, {\theta_{cav} \left(  \phi, q, k \right) \over \pi}, 
\label{fpcond}
\ee
where $\lambda (k) \!= \!2 \pi / k$ is the wavelength. Resonant modes occur at gap frequencies satisfying the usual Fabry-Perot condition $2 {\cal L} (k_m) / \lambda (k_m)\! =\! m\! \in \!\mathbb{Z}$, i.e. the previous resonance condition. This interpretation can be extended to situations where, in addition, a geometric cavity of length $L$ is inserted between the two mirrors so that the effective length becomes ${\cal L} \left(  \phi, q, k_m \right) \!= \!L \!+\! \lambda (k) \, \theta_{cav}  / (4 \pi) $, thus leading to a new set of topological gap modes \cite{supplement}. 

In summary, we have shown that topological properties of Fibonacci QP chains  encoded in Chern numbers labelling the dense set of spectral gaps,  are related to the existence of an underlying palindromic symmetry driven by a structural gauge phase $\phi$. Using a scattering approach, we have shown that these Chern numbers can be obtained either by following the behaviour of conveniently generated gap modes or by measuring scattering phases. Equivalently, finite length QP Fibonacci chains act as topological mirrors so that cavities defined by two such mirrors enclose modes whose properties are determined by topological features. This observation is of interest in different contexts such as Casimir physics. These results are readily applicable to the Aubry-Andre-Harper model. Finally, we expect some of the considerations presented in this letter to apply to QP systems in higher dimensions such as Penrose or Rauzy tilings \cite{goldman,Gambaudo}.

\begin{acknowledgments}
 Acknowledgments: This work was supported by the Israel Science Foundation Grant No.924/09. E.A. wishes to thank the College de France (Paris) where part of this work has been done with the support of a Chaire d'Etat. E.A and E.L. acknowledge fruitful discussions with G. Dunne.
\end{acknowledgments}

\section{Supplemental material}

\subsection{Notation and details on the scattering approach\label{sec:phase-shift-analysis}}

The scattering approach used in the letter is based on {[}17{]}. In
this section we provide details, specific to the calculation of phases
of the scattering matrix and spectra of the structures $P$, $\overrightarrow{F}_{N}$
and $\overleftarrow{F}_{N}$.

Each of these structures is considered as a 1D dielectric with a uniform
refractive index everywhere except for a finite spatial region, where
this index is modulated. The two possible types of layers, $\{A,B\}$, are characterised by their refractive indices $\left\{ n_{A},n_{B}\right\} $ and widths {$\left\{ d_{A},d_{B}\right\} $} such that $n_{A}d_{A}=n_{B}d_{B}\equiv nd$. The latter condition is the quarter wavelength layer setup which is known to enhances the spectral effect of quasi-periodicity. 

\begin{figure}
\includegraphics[bb=0bp 0bp 975bp 290bp,clip,width=1\columnwidth]{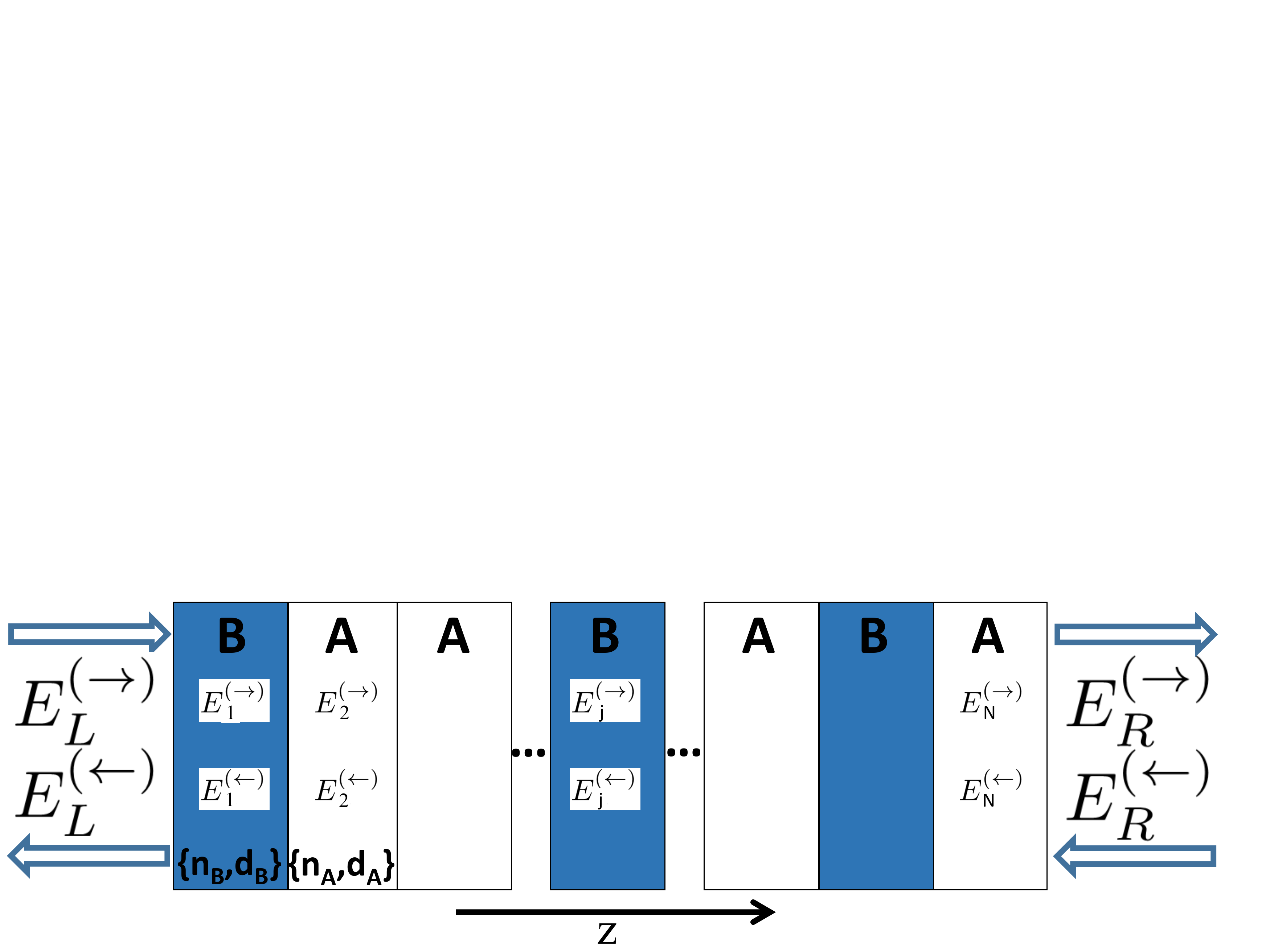}

\protect\caption{(color online) Setup and wave propagation notations for the dielectric
scattering geometry problem. $\left\{ i_{L},o_{L}\right\} \equiv\left\{ E_{L}^{(\rightarrow)},E_{L}^{(\leftarrow)}\right\} $
denote the electric field amplitudes of incoming and outgoing waves
at the left boundary, respectively. Similarly, $\left\{ i_{R},o_{R}\right\} \equiv\left\{ E_{R}^{(\leftarrow)},E_{R}^{(\rightarrow)}\right\}$ are
defined outside the right structure boundary.\label{fig:notations}}
\end{figure}

For a transverse-electric electromagnetic wave propagating in the modulation direction
$z$, the scattered quantities are electric field amplitudes, $E(z)$, such that $E(z,t)\!=\!E(z)e^{-i\omega t}$. The setup and the corresponding wave propagation notations are given in Fig.\ref{fig:notations}.

We obtain the scattering matrix defined in the letter by (4), through
the algebraically related transfer matrix, $M(k)$, which is defined by

\begin{eqnarray}
\left(\begin{array}{c}
o_{R}\\
i_{R}
\end{array}\right) & \equiv & M(k)\left(\begin{array}{c}
i_{L}\\
o_{L}
\end{array}\right)=\dfrac{1}{t}\begin{pmatrix}t/t^{*} & \overleftarrow{r}\\
-\overrightarrow{r} & 1
\end{pmatrix}\left(\begin{array}{c}
i_{L}\\
o_{L}
\end{array}\right).\label{eq:tmatrix1}
\end{eqnarray}
The transfer matrix of $\overrightarrow{F}_{N}$ is calculated
using its multiplicative properties through $M_{\overrightarrow{F}}=M_{\chi_{1}}\cdot M_{\chi_{2}}\cdot...\cdot M_{\chi_{N}}$. 

As stated in the  letter, the normalised IDOS of a structure of
length $N$ is given by $\mathcal{N}(k)=\frac{1}{\pi N}\delta(k)$.
This implies that for every new mode, the total phase shift $\delta(k)$
experiences a discrete jump of size $\pi$. This understanding will
be used in subsection \ref{sub:Resonant-conditions-for} to derive
the gap modes resonant condition.

\subsection{scattering analysis for the artificial palindrome\label{sec:phase-shift-analysis-composite}}

In this section we describe (A) the derivation of the phase $\varphi$
defined in (7) of the  letter, (B) the derivation of the resonant
condition $\theta_{cav}\left(k_{m}\right)=2\pi m$, and (C) the dependence
of this condition upon dielectric contrast.

\subsubsection{Derivation of the phase $\varphi$}

To derive the phase $\varphi$ appearing in (7) of the  letter,
one needs to relate the total phase shift of the scattering matrix
of the structure $P$, 

\begin{equation}
\begin{array}{c}
S_{P}=\left(\begin{array}{cc}
\overrightarrow{r}_{P} & t_{P}\\
t_{P} & \overleftarrow{r}_{P}
\end{array}\right)\end{array},
\end{equation}
to the total phase shift of the substructures scattering matrices  $S_{\overrightarrow{F}}$, defined
in (4) of the letter, and the corresponding $S_{\overleftarrow{F}}$.
To that purpose, we make use of the multiplicative properties of the
transfer matrix,

\begin{equation}
\begin{array}{c}
M_{P}=\left(\begin{array}{cc}
\;1/{t_P}^* & {\overleftarrow{r}_{P}}/{t_{P}}\\
{-\overrightarrow{r}_{p}}/{t_{P}} & \,1/{t_{P}}
\end{array}\right)=M_{\overrightarrow{F}}M_{\overleftarrow{F}}\end{array}.\label{eq:MP}
\end{equation}
Using (\ref{eq:tmatrix1}), and recalling that $t_{\overrightarrow{F}}=t_{\overleftarrow{F}}\equiv t$,
$M_{P}$ becomes

\begin{equation}
\begin{array}{c}
M_{P}=\dfrac{1}{t^{2}}\begin{pmatrix}t/t^{*} & \overleftarrow{r}_{\overrightarrow{F}}\\
-\overrightarrow{r}_{\overrightarrow{F}} & 1
\end{pmatrix}\begin{pmatrix}t/t^{*} & \overleftarrow{r}_{\overleftarrow{F}}\\
-\overrightarrow{r}_{\overleftarrow{F}} & 1
\end{pmatrix}\end{array}.\label{eq:MP-1}
\end{equation}
Using (6) of the letter, with (\ref{eq:MP}) and (\ref{eq:MP-1}) here, the total phase
shift $\delta_{P}$, may be extracted from $e^{2i\delta_{P}}=-t_{P}/t_{P}^{*}=\overleftarrow{r}_{P}/\overrightarrow{r}_{P}^{*}$.
Noting that $\overrightarrow{r}_{\overrightarrow{F}}=\overleftarrow{r}_{\overleftarrow{F}}\mbox{ and \ensuremath{\overrightarrow{r}_{\overleftarrow{F}}=\overleftarrow{r}_{\overrightarrow{F}}}},$
we arrive, after some straightforward algebra, to (7) of the  letter.

\subsubsection{Derivation of the resonant condition $\theta_{cav}\left(k_{m}\right)=2\pi m$
\label{sub:Resonant-conditions-for}}

The appearance of new modes in the spectrum of the structure $P$ is understood from the total phase shifts difference $\delta_{P}-2\delta$.
This results from the fact that in the ideal case where $|r|^{2}=1$,
i.e. when $\delta=const$, the appearance
of every new mode yields a discrete jump of size $\pi$ in $\delta_{P}$
(see section \ref{sec:phase-shift-analysis} above). Therefore, when
$|r|^{2}=1$, the spectral locations of new gap modes may be expressed
by the resonant condition $\delta_{P}-2\delta=\pi m$,
which according to (7) of the  letter yields $\varphi\left(k_{m}\right)=2\pi m$.

\begin{figure}
\includegraphics[bb=0bp 0bp 580bp 320bp,clip,width=1\columnwidth]{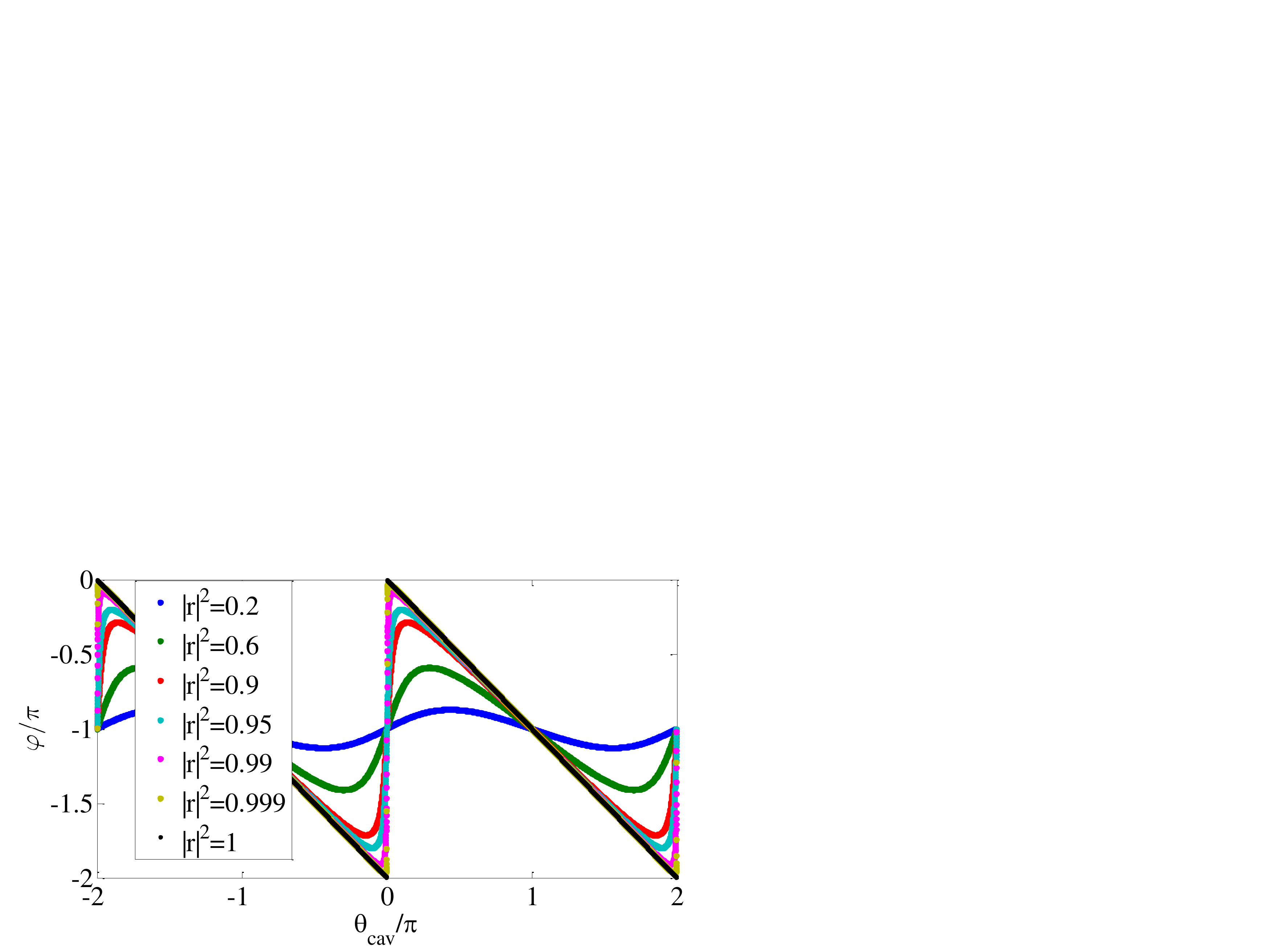}

\protect\caption{(color online) $\varphi\left(\theta_{cav},|r|^{2}\right)$ as calculated
from (\ref{eq:trigophi2}). The relation $\varphi=-\theta_{cav}$
holds only for perfect reflectance.\label{fig:modCP-1}}
\end{figure}

Next, we find the resonant condition for any $|r|^{2}\neq0$. Using (7) of the  letter, one obtains

\begin{equation}
\begin{array}{c}
|r|^{2}\cos\left(\theta_{cav}+\frac{\varphi}{2}\right)=\cos\left(\frac{\varphi}{2}\right),\end{array}\label{eq:trigophi}
\end{equation}
or alternatively,

\begin{equation}
\varphi\left(\theta_{cav},\,|r|^{2}\right){=2\mbox{Arctan}}{\left(\frac{|r|^{2}\cos \theta_{cav}-1}{|r|^{2}\sin \theta_{cav}}\right)}.\label{eq:trigophi2}
\end{equation}
From (\ref{eq:trigophi}), one can see that for perfect reflectance
$\varphi\left(\theta_{cav},\,|r|^{2}=1\right)=-\theta_{cav}$ , which
yields the condition for the appearance of new modes $\theta_{cav}\left(k_{m}\right)=2\pi m$,
as in the  letter. Using (\ref{eq:trigophi2}), Fig.\ref{fig:modCP-1}
shows the effect of varying $|r|^{2}$ on the relation between $\varphi$,
and $\theta_{cav}$: $\varphi$ no longer covers the interval $[0,2\pi]$,
and consequently never takes the perfect reflection resonant values
(while $\theta_{cav}$ does). Instead, it becomes increasingly smeared
as $|r|^{2}$ decreases.
This shows that in general, $\theta_{cav}$ is more suitable than
$\varphi$ for identifying the emergence of new modes in structures
with imperfect reflection. 

\begin{figure}
\includegraphics[bb=0bp 0bp 570bp 500bp,clip,width=1\columnwidth]{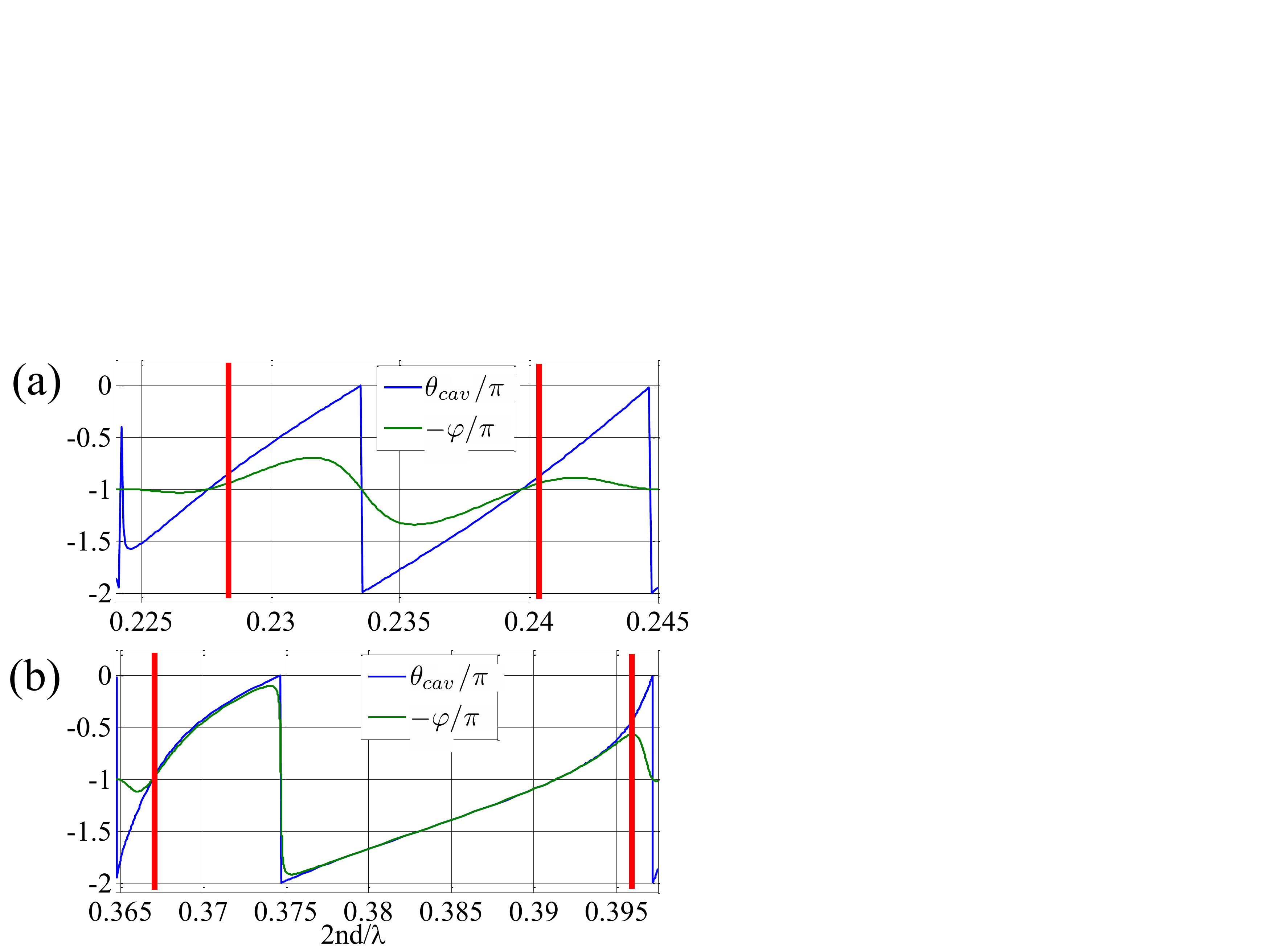}

\protect\caption{(color online) Behaviour of the phases $\theta_{cav}\left(k\right)$
and $\varphi\left(\theta_{cav}\left(k\right),|r(k)|^{2}\right)$ as a function of $k$ for the dielectric
structure {[}27{]}, for different spectral gaps (gap
edges are indicated by red bars). (a) Gap $q\!=\!2$, with $|r|^{2}\!\approx\!0.25$.
(b) Gap $q\!=\!-1$, with $|r|^{2}\!\approx\!0.98$. \label{fig:modCP-1-1}}
\end{figure}

Figure \ref{fig:modCP-1-1} shows the same
conclusion for the dielectric structure {[}27{]}. The behaviour of $\theta_{cav}\!=\!2\overrightarrow{\theta}_{\overleftarrow{F}}$,
and $\varphi\!=\!2\delta_{P}\!-\!4\delta$ is examined
for two different spectral gaps, characterised by different values
of $|r|^{2}$. Indeed, the winding range of $\theta_{cav}$ is unaffected
by $|r|^{2}$, while $\varphi$ ceases to cover the interval $[0,2\pi]$
(even for $|r|^{2}=0.98$, Fig.\ref{fig:modCP-1-1}.b). Thus, for
any value of $|r|^{2}$, the condition $\theta_{cav}\left(k_{m}\right)=2\pi m$
can be used to calculate the gap mode frequencies for $P$.

\subsubsection{Modulation strength dependence \label{sub:contrast}}

\begin{figure}
\includegraphics[bb=0bp 0bp 600bp 550bp,clip,width=1\columnwidth]{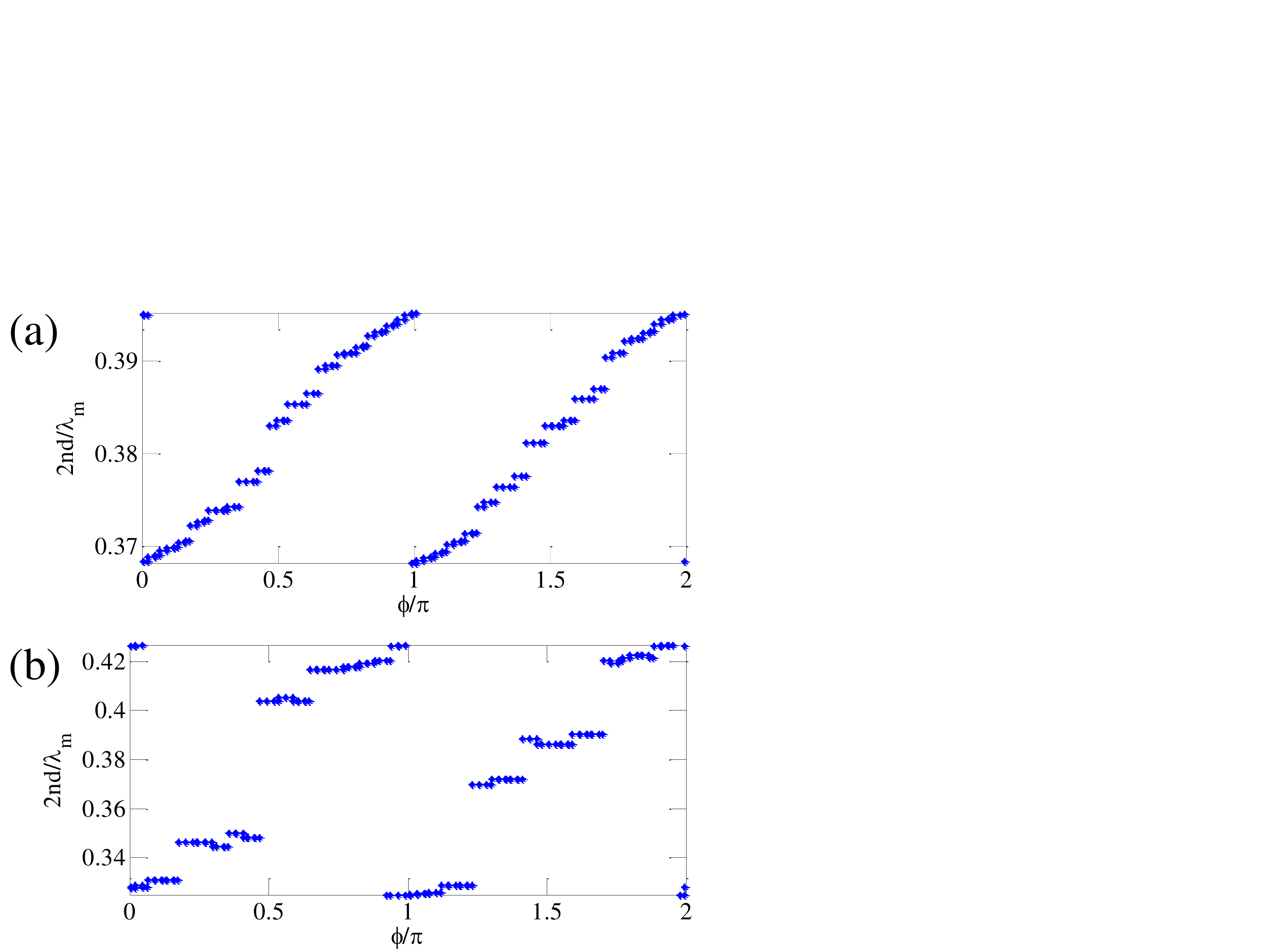}

\protect\caption{(color online) The spectral location of gap modes $k_{m}(\phi)$ as
a function of contrast for the $q=-1$ gap of the structure $P$ {[}27{]} at (a) $15\%$ and (b)
$80\%$ dielectric contrast. \label{fig:FPresonance-1}}
\end{figure}

In dielectric structures, modulation strength is given by the dielectrc contrast, $(n_{high}\!-\!n_{low})/n_{low}$. In structures with spectral gaps, such as the Fibonacci chain, increasing
the dielectric contrast results in an increase in gap widths and in
the reflectance at gap frequencies. It is thus natural to ask how
does the dielectric contrast affect the spectral topological properties
discussed in the  letter. The $\pi/|q|$ period of the gap mode crossover
(Fig.1b in the  letter), and of the $\theta_{cav}$
and $\alpha$ are found to be unchanged by the dielectric contrast.
However, the stepwise path taken by the gap modes while crossing the
gaps, and the corresponding stepwise behaviour of the phases change
significantly with the contrast. The discrete spectral changes in
the phases and in the gap mode spectral location are not of equal
magnitude as some changes are larger than others (this can be seen
for instance in Figs.5.a,d of the  letter). When the contrast is increased,
the larger spectral jumps increase, while smaller spectral jumps decrease.
For too large dielectric contrast, the entire crossover is accomplished
by one spectral jump, thus unmeasurable. This trend can be seen in
Fig.\ref{fig:FPresonance-1}.

\subsection{Effective Fabry-Perot model with a cavity\label{sec:The-generalized-fabry-perot}}

\begin{figure}
\includegraphics[bb=0bp 0bp 600bp 650bp,clip,width=1\columnwidth]{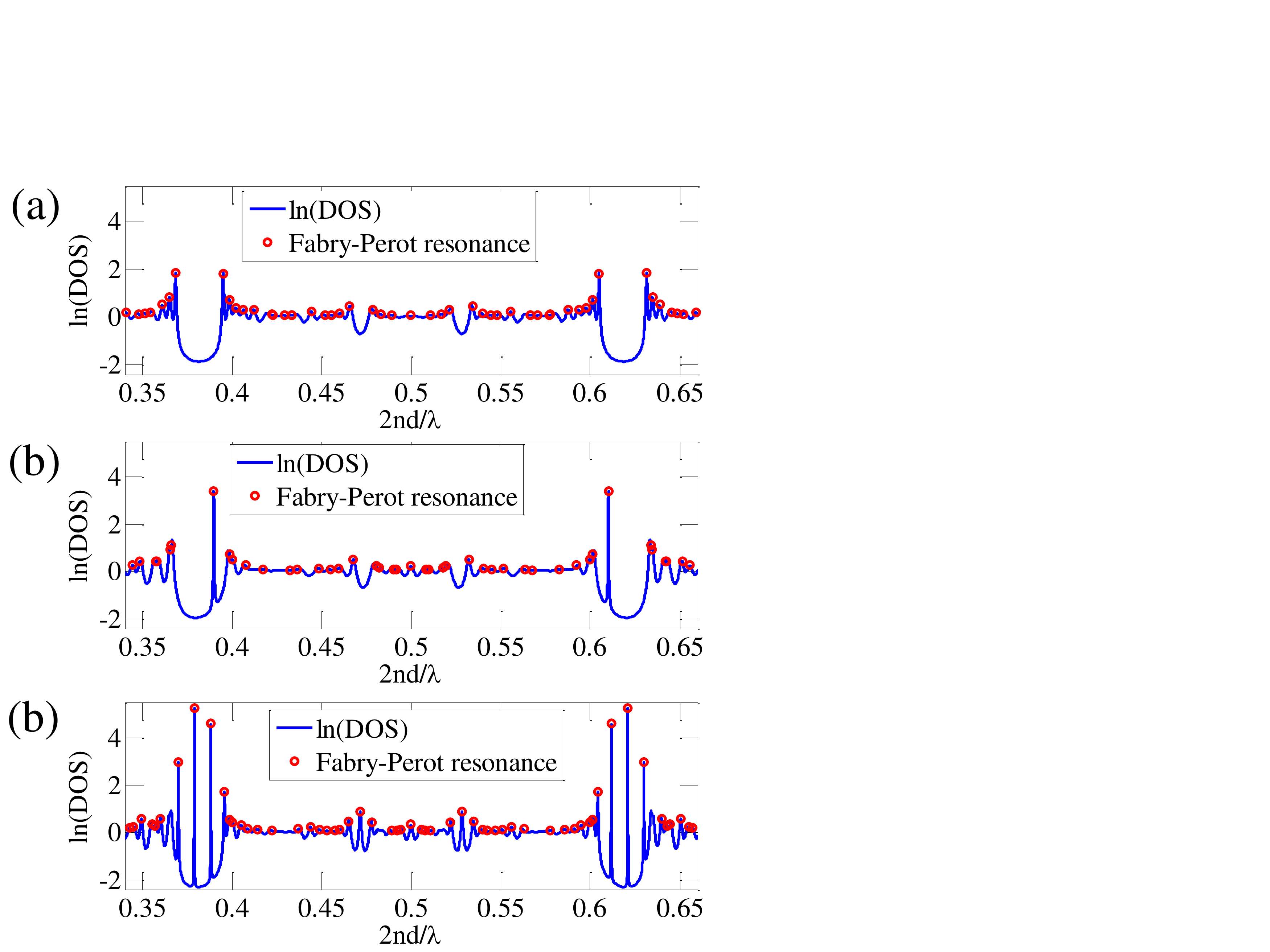}

\protect\caption{(color online) (a)-(b) Density of states for the structure
$P$ {[}27{]} (blue line). Red circles mark the Fabry-Perot
resonant wave-vector values $k_{m}$, obeying $\theta_{cav}\left(k_{m}\right)\!=\!2\pi m$. (a) $\phi\!=\!0$, yielding a palindromic structure. (b) A non-palindromic structure ($\phi=1.7\pi$). (c) The
genuine cavity setup: the substructures $\protect\overrightarrow{F}_{N}\mbox{, and }\protect\overleftarrow{F}_{N}$
are separated by a uniform region with a refractive index $n\!=\!n_{A}$,
and width $d_{free}\!=\!80nd$. The density of states (blue lines) and
the resonant condition prediction (red circles) are given for $\phi=1.7\pi$.
\label{fig:FPresonance}}
\end{figure}

In this section, we generalise the effective Fabry-Perot interpretation
to include the case of a genuine cavity inserted between the topological
mirrors discussed in the  letter. In the standard description, a Fabry-Perot cavity is defined by two mirrors separated by the length $L,$, and a discrete spectrum obtained by the resonant
condition $2L/\lambda_{m}=m$, may be viewed as the winding of
a (frequency dependent) cavity phase, $\theta_{L}\equiv4\pi L/\lambda$. This
leads to a resonance condition $\theta_{L}\left(k_{m}\right)=2\pi m$.
In the effective Fabry-Perot model for the structure $P$,
waves are also localized inside a finite region (the interface region)
by means of distributed feedback, without a geometrical
cavity, but only the sum of the two mirrors reflected phase shifts,
$\theta_{cav}$. Substituting $\theta_{L}$ by $\theta_{cav}$ yields
the definition of the virtual (frequency dependent) cavity length
$\mathcal{L}$, given by (11) of the  letter. Figures \ref{fig:FPresonance}.a
and \ref{fig:FPresonance}.b display the predictions of the resonant condition
for palindromic and non-palindromic values of $\phi$,
respectively.

As discussed in the  letter, the effective Fabry-Perot model may be
generalised to the case where in addition to the previous case, a genuine cavity of length $L$ with a uniform
refractive index, is inserted in between the topological mirrors. This
genuine cavity should be added to
the effective one. In this case, the resonances may be found in the
cavity phase picture as

\begin{equation}
\theta_{total}\left(k_{m}\right)=\theta_{cav}\left(k_{m}\right)+\theta_{L}\left(k_{m}\right)=2\pi m,
\end{equation}
or alternatively in the cavity length picture given by

\begin{equation}
2L_{total}\left(k_{m}\right)/\lambda\left(k_{m}\right)=m,
\end{equation}
with $L_{total}=L+\mathcal{L}$. As in the standard Fabry-Perot cavity case, the increase in the effective
cavity length reduces the resonant mode separation. Since the gap
widths did not change, multi-resonant-modes appear now in the gaps (Fig.\ref{fig:FPresonance}.c).

\subsection{The gap labeling theorem\label{sec:GLT}}

In the following section, we present an elementary derivation of (1) in the  letter, following the gap labeling theorem {[}1{]} which applies to quasicrystals which can be constructed
by substitution rules. Considering a 2 letter alphabet $\{A\:, B\}$,
and the substitutions $\{\sigma(A),\:\sigma(B)\}$ (given in the letter), we first we define
the one and two letters occurrence matrices, $M$ and $M_{2}$ respectively.
The matrix element $M^{ij}$ contains the occurrence of the $i^{th}$
letter in the $j^{th}$ substitution. For the Fibonacci chain, it is

\begin{equation}
M=\left(\begin{array}{cc}
1 & 1\\
1 & 0
\end{array}\right).
\end{equation}

The matrix element $M_{2}^{ij}$contains the occurrence of the $i^{th}$
2-letter combination (doublet) in the $j^{th}$ 2-letter substitution.
For the Fibonacci chain, the 3 doublets alphabet is $\{AA,\, AB,\, BA\}$, and the corresponding substitutions are $\{\sigma(AA)=ABAB,\,\sigma(AB)=ABA,\,\sigma(BA)=AAB\}$. No $BB$ doublets occur. The occurrences of the doublets in the 2-letter substitutions are obtained as follows {[}1{]}. The numbers of doublets to be deduced
(by shifting a 2-letter acceptance window from left to right) is equal
to the number of letters in the substitution for the single first
letter of the doublet. For $\sigma(AA)$ and $\sigma(AB)$ we
should deduce 2 doublets (as the length of $\sigma(A)=AB$ is
2), and for $\sigma(BA)$ we should deduce $1$ doublet (as the
length of $\sigma(B)=A$ is $1$). Finally,

\begin{equation}
\sigma_{2}(AA)=\sigma_{2}(AB)=(AB)(BA)\:;\:\sigma_{2}(BA)=(AA),
\end{equation}
which leads to the occurrence matrix
\begin{equation}
M_{2}=\left(\begin{array}{ccc}
0 & 0 & 1\\
1 & 1 & 0\\
1 & 1 & 0
\end{array}\right).
\end{equation}
The next step is to diagonalize $M$ and $M_{2}$, and normalize the
eigenvectors corresponding to the highest mutual eigenvalue, so that the sum of their elements is unity. 
The eigenvalues of $M$ and $M_{2}$ are $\{1\!-\!\tau,\!\tau\}$ and $\{1\!-\!\tau,\!\tau,\!0\}$, respectively.
Thus,  $\tau$ is the mutual highest eigenvalue, and the corresponding normalized
eigenvectors $V_{1},\: V_{2}$ are

\begin{equation}
V_{1}=\left(\begin{array}{c}
\tau^{-1}\\
1-\tau^{-1}
\end{array}\right);V_{2}=\left(\begin{array}{c}
2\tau^{-1}-1\\
1-\tau^{-1}\\
1-\tau^{-1}
\end{array}\right).
\end{equation}
Using the identity $\tau^{-1}=\tau-1$, (12) becomes
\begin{equation}
V_{1}=\left(\begin{array}{c}
\tau-1\\
2-\tau
\end{array}\right);V_{2}=\left(\begin{array}{c}
2\tau-3\\
2-\tau\\
2-\tau
\end{array}\right).
\end{equation}

The elements $V_{1}^{(i)}$ ($V_{2}^{(i)}$) represent the occurrence
of the $i^{th}$ letter(doublet). From $V_{1}$ we see that
the letter $A$ occurs more than the letter $B$ by $61.8\%:38.2\%$.
From $V_{2}$ we see that the $AB$, and $BA$ doublets have an
equal occurrence of $\sim\!38.2\%$
each, and the $AA$ doublet has $\sim\!23.6\%$ occurrence. This
clarifies the asymptotic value of the plateau in $\eta(\phi)$  (Fig.2.d in the  letter). For the
$\sim\!23.6\%$ $AA$ doublet occurrence in a given structure length there
can never be a corresponding $BB$ doublet preventing $\eta(\phi)$
from approaching unity (anti-palindrome).

The normalized IDOS in the gaps
can be calculated as follows {[}1{]}. Let $\mathbb{Z}[x]$
be the set of finite polynomials of degree $N$ with integer coefficients,
\begin{equation}
\mathbb{Z}[x]=\left\{ \left. \sum_{n=1}^{N}q_{n}x^{n}\right|q \in\mathbb{Z},\, N\in\mathbb{N}\right\} .
\end{equation}
Now, $\mathcal{N}_{gap}$, the IDOS at possible gaps is equal to {[}1{]}
\begin{equation}
\mathcal{N}_{gap}=\left\{ \left.a\boldsymbol{P}\left(\vartheta^{-1}\right)\right|\boldsymbol{P}\left(\vartheta^{-1}\right)\in\mathbb{Z}\left[\vartheta^{-1}\right]\right\} \: mod.\:1,
\end{equation} 
where $a\in\{V_{1},V_{2}\}$, and $\vartheta$ is the highest common
eigenvalue of the occurrence matrices (i.e. $\vartheta=\tau$
for the Fibonacci chain). Using (12), the set appearing in (15) for
the Fibonacci chain is reduced to
\begin{eqnarray}
\mathcal{N}_{gap} & = & \left\{ q_{1}\frac{\tau^{-1}}{\tau^{n_{1}}};q_{2}\frac{\left(1\!-\!\tau^{-1}\right)}{\tau^{n_{2}}};q_{3}\frac{\left(2\tau^{-1}\!-\!1\right)}{\tau^{n_{3}}}\right\} \nonumber \\
 &  & mod.\:1,
\end{eqnarray}
 where $n_{1,2,3}\!\in\!\mathbb{N}$, and $q_{1,2,3}\!\in\!\mathbb{Z}$. Since
$\tau^{-2}\!=\!1\!-\!\tau^{-1}$, the set can be further compacted into
\begin{equation}
\mathcal{N}_{gap}=\left\{ q\tau^{-1},q\in\mathbb{Z}\right\} \: mod.\:1,
\end{equation}
where $q\in\mathbb{Z}$. This form leads to (1) of the letter.


\begin{thebibliography}{99}

\bibitem{gaplabel} J. Bellissard, A. Bovier and J.M. Ghez, Reviews in Math. Physics, Vol. 4, No. 1, 1-37 (1992) and J. Bellissard, Les Houches, Springer, J.M. Luck, P. Moussa and M. Waldschmidt Eds., (1993), S. Ostlund, R. Pandit, D. Rand, H.J. Schellnhuber and E.D. Siggia,Phys. Rev. Lett. \textbf{50}, 1873 (1983) and S. Ostlund, D. Rand, J. Sethna and E.D. Siggia, Physica 8D, 303 (1997).

\bibitem{general} B. Simon, Adv. Appli. Math. {\bf 3}, 463 (1982), M. Shubin, Russian Math. Surveys   \textbf{34}, 109 (1979), R. Johnson and J. Moser, Commun. Math. Phys. 84, 403 (1982), R. Johnson, Acta Applicandae Mathematicae, {\bf 1}, 241 (1983) and K. Bjerklov,  T. Jager, J. of American. Math. Soc. {\bf 22}, 353 (2009), J. Kellendonk, Commun. Math. Phys. 187, 115 (1982) and D. lenz and R.V. Moody, Commun. Math. Phys. 289, 907 (2009).


\bibitem{general2} M. Kohmoto, L. P. Kadanoff, and C. Tang, Phys. Rev. Lett. 50, 1870 (1983); S. Ostlund and S. Kim, Phys. Scr. T9, 193 (1985), E.L. Albuquerque and M.G. Cottam, Phys. Rep. 376, 225 (2003).

\bibitem{hof} D.R. Hofstadter, Phys. Rev. B {\bf 14}, 2239 (1976).

\bibitem{aubry} S. Aubry and G. Andre, Ann. Israel Phys. Soc. {\bf 3}, 134 (1980).

\bibitem {damanik}D. Damanik and A. Gorodetski, Commun. Math. Phys.
\textbf{305}, 221 (2011) and D. Damanik, M. Embree, A. Gorodetski, S. Tcheremchantsev,  Commun. Math. Phys. \textbf{280}, 499 (2008)

\bibitem{spontem} E. Akkermans and E. Gurevich, Europhys. Lett. \textbf{103}, 30009 (2013).

\bibitem{reviewfractals} For a recent review, E. Akkermans, Contemporary Mathematics {\bf 601}, 1-22 (2013), arXiv:1210.6763.

\bibitem{kunz} H. Kunz, Phys. Rev. Lett. 57, 1095 (1986)

\bibitem{polaritons} D. Tanese, E. Gurevich, F. Baboux, T. Jacqmin, A. Lema\^{\i}tre, E. Galopin, I. Sagnes, A. Amo, J. Bloch and E. Akkermans, Phys. Rev. Lett. \textbf{112}, 146404 (2014) 

\bibitem{krauss2} Y.E. Kraus, Y. Lahini, Z. Ringel, M. Verbin and O. Zilberberg, Phys. Rev. Lett. \textbf{109}, 106402 (2012)

\bibitem{dana} I. Dana, Phys. Rev. B \textbf{89}, 205111 (2014)

\bibitem{substitution} M. Duneau and A. Katz, Phys.Rev.Lett. {\bf 54}, 2688 (1985), P.A. Kalugin, A. Kitaev and L. Levitov, JETP Lett. {\bf 41}, 145 (1985), J. Physique Lett. (Paris) {\bf 46}, L601 (1985) and M. Duneau, R. Mosseri and C. Oguey, J. Phys. A  {\bf 22}, 4549 (1989) 

\bibitem{grimm} M. Baake and U. Grimm, "Aperiodic Order", Volume 1: A Mathematical Invitation, The Open University, Milton Keynes, E. Bombieri and J.E. Taylor, J. Physique \textbf{C3} , 19 (1986), V. Elser, Acta Cryst. A {\bf 42}, 36 (1986) and J.M.Luck, C. Godreche, A. Janner and T.Janssen, J. Phys. A {\bf 26}, 1951 (1993).

\bibitem{supplement} See Supplemental Material.

\bibitem{luck} H. Kesten, Acta Arith. {\bf 12}, 193 (1966), J. M. Luck, Phys. Rev. B \textbf{39}, 5834 (1989), S. Ostlund and R. Pandit, Phys. Rev. B \textbf{29}, 1394 (1984) and C. Godreche and C. Oguey, J. Physique (France) {\bf 51}, 21 (1990).

\bibitem{krauss} Y.E. Kraus and O. Zilberberg, Phys. Rev. Lett. \textbf{109}, 116404 (2012)

\bibitem{meidan} D. Meidan, T. Micklitz and P. W. Brouwer, Phys. Rev. B \textbf{84}, 195410-1 (2011) and  I. C. Fulga, F. Hassler, and A. R. Akhmerov, Phys. Rev. B \textbf{85}, 165409 (2012).

\bibitem{dunnelevyea} E. Akkermans, G. V. Dunne, and E. Levy, 
in "Optics of Aperiodic Structures: Fundamentals and  Device Applications", L. dal Negro ed., Pan Stanford Publishing, (2013)

\bibitem{rkphidep} A remaining $\phi$-dependence, apparent in Fig.\ref{spectral}.c-d, is irrelevant to the results of this paper \cite{unpub}.

\bibitem{rkcredit}  The emergence of boundary gap modes is  well known in photonic crystals \cite{gapmodpc} and in Fibonacci chains \cite{gapmodfibo}. Recently, gap modes in the Aubry-Andre-Harper tight binding model \cite{aubry} together with their $\phi$ driven topological features have been addressed \cite{krauss2}.

\bibitem{unpub} E. Levy and E. Akkermans (unpublished).

\bibitem{rkdoubling} The absence of boundary conditions for the open structure $\overrightarrow{F}_N \overleftarrow{F}_N$  implies that its inner boundary involves twice as many gap modes than a reflective boundary. The $\pi/|q|$ gap mode crossover period is observed only through this approach.

\bibitem{rklevinson} This condition can be seen as a resonance condition obtained from the Levinson theorem \cite{ma}. 

\bibitem{rkalphak} A residual and monotonic $k$-dependence within the gaps is apparent in  Figs.\ref{chern11}.a,e. But, it does not affect the topological properties \cite{unpub}.

\bibitem{rkhafezi} Note the factor 2 between the phase shift formula for 
  the Chern number in \cite{dana,hafezi} and in (\ref{winding}). It is characteristic of the $\pi$-periodicity of the palindromic cycle \cite{rkdoubling}.

\bibitem{goldman} D. T. Tran, A. Dauphin, N. Goldman and P. Gaspard, Phys. Rev. B \textbf{91}, 085125 (2015).

\bibitem{Gambaudo} J. M. Gambaudo and P. Vignolo, New J. Phys. 16 (2014) 043013 and P. Vignolo, M. Bellec, J. Boehm, A. Camara, J.M. Gambaudo, U. Kuhl and  F. Mortessagne, arXiv:1411.1234 

\bibitem{rkstructure} The dielectric chains used throughout the letter are based on  $\overrightarrow{F}_N$ with $N= 89$, which is identical to $S_{10}$ for $\phi=\phi_F$. The letters $\{A,B\}$ are represented by the refractive indices $\{n_A,n_B\}$, and layer thicknesses $\{d_A,d_B\}$. The dielectric contrast defined as $(n_{high}-n_{low})/n_{low}$ is taken to $15\%$. The vacuum wave vector is expressed in units of $ndk/\pi = 2nd/\lambda$, where $nd=n_Ad_A$. See  \cite{supplement} for details.

\bibitem{rkspecpal}  The phase $\alpha(\phi)$ describes both spectral palindromes and anti-palindromes, and unlike $\eta (\phi)$, does not saturate.

\bibitem{ma} Zhong-Qi Ma, J. Phys. A \textbf{39}, R625 (2006) and J. Kellendonk and S. Richard, J. Phys. A . \textbf{39}, 14397 (2006).


\bibitem{hafezi} A. V Poshakinskiy, A. N. Poddubny, and M. Hafezi, Phys. Rev. A \textbf{91}, 043830 (2015).

\bibitem{gapmodpc} For a review on defect and surface modes in photonic crystals, see J. D. Joannopoulos, S. G. Johnson, J. N. Winn, and R. D. Meade, "Photonic Crystals: Molding the Flow of Light", (Princeton university press, 2011).

\bibitem{gapmodfibo}  See for instance E.S. Zijlstra, A. Fasolino, and T. Janssen, Phys. Rev. B \textbf{59}, 302 (1999), Y. El Hassouani, H. Aynaou, E. H. El Boudouti, B. Djafari-Rouhani, A. Akjouj, and V. R. Velasco, Phys. Rev. B \textbf{74}, 035314 (2006), and E. Abdel-Rahman and A. Shaarawi, J. Mater. Sci. Mater. Electron. \textbf{20}, 153 (2009).

\end{thebibliography}
\end{document}